\documentclass[aps,prd,twocolumn,nofootinbib]{revtex4}
\usepackage{graphicx}
\usepackage{epsfig,graphics,graphicx}

\newcommand{\bel}[1]{\begin{equation}\label{#1}}
\newcommand{\bal}[1]{\begin{eqnarray}\label{#1}}

\newcommand{\be}{\begin{equation}}
\newcommand{\ee}{\end{equation}}
\newcommand{\ba}{\begin{eqnarray}}
\newcommand{\ea}{\end{eqnarray}}

\newcommand{\bes}{\begin{equation*}}
\newcommand{\ees}{\end{equation*}}

\newcommand{\sig}{\langle\sigma\rangle}
\newcommand{\et}{\langle\eta\rangle}
\newcommand{\al}{\vec\alpha_0}

\begin{document}

\title{CP violation and chiral symmetry restoration in the hot linear sigma model
in a strong magnetic background}

\author{Ana J\'ulia {\sc Mizher}\footnote{anajulia@if.ufrj.br} and
Eduardo S. {\sc Fraga}\footnote{fraga@if.ufrj.br}}

\affiliation{Instituto de F\'\i sica, Universidade Federal do Rio de Janeiro, \\
Caixa Postal 68528, Rio de Janeiro, RJ 21941-972, Brazil}

\begin{abstract}
We study the effects of CP violation on the nature of the chiral transition within
the linear sigma model with two flavors of quarks. The finite-temperature effective
potential containing contributions from nontrivial values for the parameter $\theta$
is computed to one loop order and their minima structure is analyzed. Motivated by
the possibility of observing the formation of CP-odd domains in high-energy heavy
ion collisions, we also investigate the behavior of the effective potential in the presence
of a strong magnetic background. We find that the nature of the chiral transition is
influenced by both $\theta$ and the magnetic field.
\end{abstract}

\maketitle

\section{Introduction}

In Quantum Chromodynamics (QCD), the existence of instanton configurations for the gauge
fields \cite{Belavin:1975fg} and their intimate connection with the axial anomaly \cite{ABJ}
allows for a nontrivial topological term in the action. This term is usually neglected, since it
breaks the original CP symmetry and so far experiments indicate that its coefficient,
known as the $\theta$ parameter, is vanishingly small, $\theta \lesssim 10^{-10}$ \cite{exp-theta}
(see also \cite{neutron-dipole}).
The reason why $\theta$ is so small (or zero) is unclear, and this issue is called the
{\it strong CP problem} \footnote{Frameworks that preserve the CP symmetry were also
proposed, the most studied being axion theories \cite{axions}, but those were
not confirmed by experiments to date either, leaving the problem unsolved.}.

Although it has been proved that CP can not be spontaneously broken in the vacuum
of  QCD for $\theta =0$ \cite{Vafa:1984xg}, this theorem might not hold for QCD matter at
finite temperature or density \cite{finitetemp}. With this premise, it has been proposed
in Ref. \cite{Kharzeev:1998kz} that hot matter produced in ultra-relativistic heavy ion collisions
could exhibit domains of metastable states that violate CP. These states could be
described by a QCD action that incorporates the topological $\theta$-term, and would decay
via CP-odd processes. Possible experimental signatures for the presence of CP-odd
domains are based on charge separation of hadronic matter produced in heavy ion
collisions \cite{Kharzeev:1999cz}. This effect is enhanced by the presence of a strong magnetic
background in the case of noncentral collisions, as was realized more recently, by a
mechanism that has been called {\it chiral magnetic effect} \cite{Kharzeev:2007jp}, and
could in principle be observed a RHIC and the LHC.

In this paper we investigate how the chiral transition is affected when the ingredients mentioned
above, i.e. CP-odd effects and a strong magnetic background, are present, as in the case of
noncentral heavy-ion collisions. For this purpose, we adopt the linear sigma model coupled with
two flavors of quarks as our effective theory \cite{GellMann:1960np} to study the chiral transition
\footnote{This effective theory, especially the $\sigma$-$\pi$-quark sector, has been widely used
to describe different aspects of the chiral transition, such as thermodynamic
properties \cite{quarks-chiral,ove,Scavenius:1999zc,Caldas:2000ic,Scavenius:2000qd,
Scavenius:2001bb,paech,Mocsy:2004ab,Aguiar:2003pp,Schaefer:2006ds,Taketani:2006zg}
and the nonequilibrium phase conversion process \cite{Fraga:2004hp},
as well as combined to other models in order to include effects from
confinement \cite{polyakov,explosive}, usually without the inclusion of the $\theta$ term.}.
To include the effects of the presence of the axial anomaly and CP violation, we add a term
that mimics the presence of nontrivial gauge field configurations, the 't Hooft
determinant \cite{'tHooft:1986nc}. The latter is a function of the parameter $\theta$,  and is
responsible for CP violation for non-vanishing values of $\theta$. In this framework, the
presence of CP violation is directly related to a nonzero $\eta$ condensate.

The rich vacuum structure brought about by a nonzero $\theta$ term in the
action \cite{Callan:1976je} has an influence on the chiral transition, and generates a more
complex picture in the analysis of the phase diagram of strong interactions
\footnote{See, for instance, Ref. \cite{Boer:2008ct} for a detailed study of the phase
structure of the two-flavor NJL model at $\theta =\pi$.}. As will be shown below,
three phases emerge: one with $\sigma$ and $\eta$ condensates, another where the $\eta$
condensate vanishes and the $\sigma$ condensate remains, and a phase where both condensates
are melted. The topography of extrema is therefore rich, and metastable minima do appear in certain
situations.

Once we have an effective theory describing chiral symmetry restoration at finite temperature
in the presence of CP violation, we study how this CP-odd linear sigma model is affected
by the presence of a strong magnetic background, as is presumably generated in the case
of noncentral high-energy heavy ion collisions \cite{Kharzeev:2007jp}. This extends the analysis
we presented in Ref. \cite{Fraga:2008qn}, where we investigated the effects of a strong magnetic
background on the nature and dynamics of the chiral phase transition at finite temperature and
vanishing chemical potential, and found that the nature of the chiral transition is modified. Here
we find that the main result obtained in \cite{Fraga:2008qn}, namely that the presence of a strong
magnetic background turns the crossover into a weak first-order phase transition, is maintained.
In fact, the combination of effects from high magnetic fields and the CP-odd contribution allows
for nontrivial combinations of metastable CP-odd and chirally symmetric phases, a structure
that might be relevant for the supercooling dynamics that presumably happens after a high-energy
heavy ion collision.

Modifications in the vacuum of CP-symmetric QCD by the presence of a magnetic field
have been investigated previously within different frameworks, mainly using effective
models \cite{Klevansky:1989vi,Gusynin:1994xp,Babansky:1997zh,Klimenko:1998su,
Semenoff:1999xv,Goyal:1999ye,Hiller:2008eh,Rojas:2008sg},
especially the NJL model \cite{Klevansky:1992qe}, and chiral perturbation
theory \cite{Shushpanov:1997sf,Agasian:1999sx,Cohen:2007bt}, but also resorting to the
quark model \cite{Kabat:2002er} and certain limits of QCD \cite{Miransky:2002rp}.
Most treatments have been concerned with vacuum modifications by the magnetic field,
though medium effects were considered in a few cases, as e.g. in the study of the stability
of quark droplets under the influence of a magnetic field at finite density and zero
temperature, with nontrivial effects on the order of the chiral transition \cite{Ebert:2003yk}.
More recently, magnetic effects on the dynamical quark mass \cite{Klimenko:2008mg} and on
the thermal quark-hadron transition \cite{Agasian:2008tb}, as well as  magnetized chiral
condensates in a holographic description of chiral symmetry breaking \cite{holographic}, were
also considered.

The paper is organized  as follows. Section II presents the low-energy effective model adopted
in this paper to investigate the chiral transition in the presence of CP violating terms. In Section III
we show our results for the effective potential at finite temperature, and discuss the condensates.
In section IV we incorporate effects coming from the presence of a strong magnetic background.
Section $V$ contains our conclusions and outlook.

\section{CP-odd linear sigma model}

The classical Lagrangian for QCD is invariant under parity, $P$, and charge conjugation, $C$,
transformations. However, at a quantum level these symmetries are not preserved. This breaking
manifests itself as the axial (or ABJ) anomaly \cite{ABJ}, and the current associated with this symmetry
is no longer conserved, even in the massless limit for quarks:
\be
\partial_\mu J^{\mu}_{5} = 2m_f i\bar\psi_f\gamma_5\psi_f -
\frac{N_f g^2}{16\pi^2}F^{\mu\nu a} \tilde F^{a}_{\mu\nu} \; ,
\ee
where g is the gauge coupling, $N_{f}$ is the number of quark flavors with masses $m_{f}$,
$\tilde F$ is the dual of the gauge field strength tensor $F$, and sums over color indices $a$
and flavor indices $f$ are implicit.
Naively, the last term in the expression above would be irrelevant, since it can be written as a total
derivative. However, if one considers topologically nontrivial gauge field configurations, this term is
nonvanishing and can be associated with the winding number of each configuration. Each sector of configurations, with a given winding number, has its own vacuum, and the true vacuum of the theory
becomes a superposition of the several inequivalent topological vacua
$|\theta\rangle = \sum_n e^{-in\theta}|n\rangle$, where $\theta$ is a free parameter.
In this context, to calculate expectation values one has to compute an average over a sector with
fixed winding number and then sum over all sectors. This procedure is equivalent to adding a
topological term in the Lagrangian, as follows:
\be
\mathcal{L}_{QCD} =  \mathcal{L}_{cl}
- \frac{\theta}{32\pi^2} g^2 F^{\mu\nu a} \tilde F^{a}_{\mu\nu} \; .
\ee

To describe the chiral phase structure of strong interactions including CP-odd effects,
we adopt an effective model that reproduces the symmetries of QCD at low energy scales,
and has the appropriate degrees of freedom at each scale: the CP-odd linear sigma
model coupled with two flavors of quarks. The chiral mesonic sector is built including all
Lorentz invariant terms allowed by symmetry and renormalizability. Following
Refs. \cite{'tHooft:1986nc,sigma-model}, one can write
\ba\nonumber
\mathcal{L}_{\chi}&=& \frac{1}{2} {\rm Tr}(\partial_\mu\phi^\dagger \partial^\mu \phi)
+ \frac{a}{2} {\rm Tr}(\phi^\dagger \phi)
- \frac{\lambda_1}{4} [{\rm Tr}(\phi^\dagger \phi)]^2 \\ \nonumber
&&-\frac{\lambda_2}{4} {\rm Tr}[(\phi^\dagger \phi)^2]
+ \frac{c}{2}[e^{i\theta}\det(\phi) + e^{-i\theta}\det(\phi^\dagger)] \\
&&+ {\rm Tr}[h(\phi +\phi^\dagger)] \; .
\ea
The potential in the Lagrangian above displays both spontaneous and explicit symmetry breaking,
the latter being implemented by the term $\sim h$. The strength of CP violation is contained in
the 't Hooft determinant term, which encodes the Levi-Civita structure of the axial anomaly and
depends on the value of the parameter $\theta$.

Expressing the chiral field $\phi$ as
\be
\phi = \frac{1}{\sqrt{2}}(\sigma + i\eta) +
\frac{1}{\sqrt{2}} (\vec{\alpha} + i \vec{\pi}) \cdot \vec{\tau} \; ,
\ee
the potential takes the following form (substituting the parameter $h$ by $H\equiv\sqrt{2} h$):
\ba \nonumber
V_{\chi}&=& -\frac{a}{2} (\sigma^2 +\vec\pi^2 + \eta^2 + \vec\alpha_0^2)  \\ \nonumber
&&-\frac{c}{2} \cos\theta ~(\sigma^2  +\vec\pi^2 + \eta^2 + \vec\alpha_0^2) \\ \nonumber
&&+ c ~\sin\theta ~(\sigma\eta - \vec\pi \cdot\vec\alpha_0) - H\sigma \\ \nonumber
&&+ \frac{1}{4}(\lambda_1 + \frac{\lambda_2}{2}) (\sigma^2 + \eta^2 + \vec\pi^2 + \vec\alpha_0^2)^2 \\
&&+ \frac{2\lambda_2}{4}(\sigma\vec\alpha_0 + \eta\vec\pi + \vec \pi \times\vec\alpha_0)^2 \; .
\ea
The parameters $a$, $c$, $H$, $\lambda_{1}$ and $\lambda_{2}$ can be fixed by vacuum
properties of the mesons, so that the effective model reproduces correctly the phenomenology
of QCD at low energies and in the vacuum, for vanishing $\theta$, such as
the spontaneous (and small explicit) breaking of chiral symmetry and experimentally measured
meson masses.

In our treatment, the chiral mesons are coupled in the standard Yukawa fashion to two flavors of
quarks. The latter constitute a thermalized fluid that provides a background in which the long
wavelength modes of the chiral fields evolve. The full Lagrangian has the form:
\ba \nonumber
\mathcal{L} &=& \frac{1}{2}(\partial_\mu\sigma)^2
 +\frac{1}{2}(\partial_\mu\vec\pi)^2 + \frac{1}{2}(\partial_\mu\eta)^2
 + \frac{1}{2}(\partial_\mu\al)^2  \nonumber \\
&-& V(\sigma,\eta,\vec\pi,\al)
+ \bar\psi_{f}(i\gamma^\mu\partial_{\mu})\psi_{f} \nonumber \\
&-&g\bar\psi_{f}(-i\gamma^5\vec\tau\cdot\vec\pi +\sigma
 \gamma^5\vec\tau\cdot\al - i\gamma^5\eta)\psi_{f} \; .
\ea
For $T=0$ quarks degrees of freedom are absent (excited only for $T > 0$), and results from
chiral perturbation theory for the broken phase vacuum are reproduced \cite{FTFT-books}.
For $T > 0$, quarks are relevant (fast) degrees of freedom and chiral symmetry is approximately
restored in the plasma for high enough $T$. In this case, we incorporate quark thermal fluctuations
in the effective potential for the mesonic sector, i.e. we integrate over quarks to one loop.

To one loop, and within a classical approximation for the chiral field background, a standard
functional integration over the fermions \cite{FTFT-books} gives an effective potential of the
form $V_{eff} = V(\phi) + V_q(\phi)$, where the quark contribution is given by
\begin{equation}
V_q\equiv -\nu_q T \int \frac{d^3k}{(2\pi)^3} \ln\left( 1 + e^{-E_k(\phi)/T} \right) \; .
\end{equation}
where $\nu_q=24$ is the color-spin-isospin-baryon charge degeneracy factor, and
$E_k(\phi)=(\vec{k}^2+M_{q}^2(\phi))^{1/2}$, $M_{q}$ playing the role of an effective
mass for the quarks.

\begin{figure}[!ht]
\begin{center}
\begin{tabular}{ccc}
 \epsfig{file=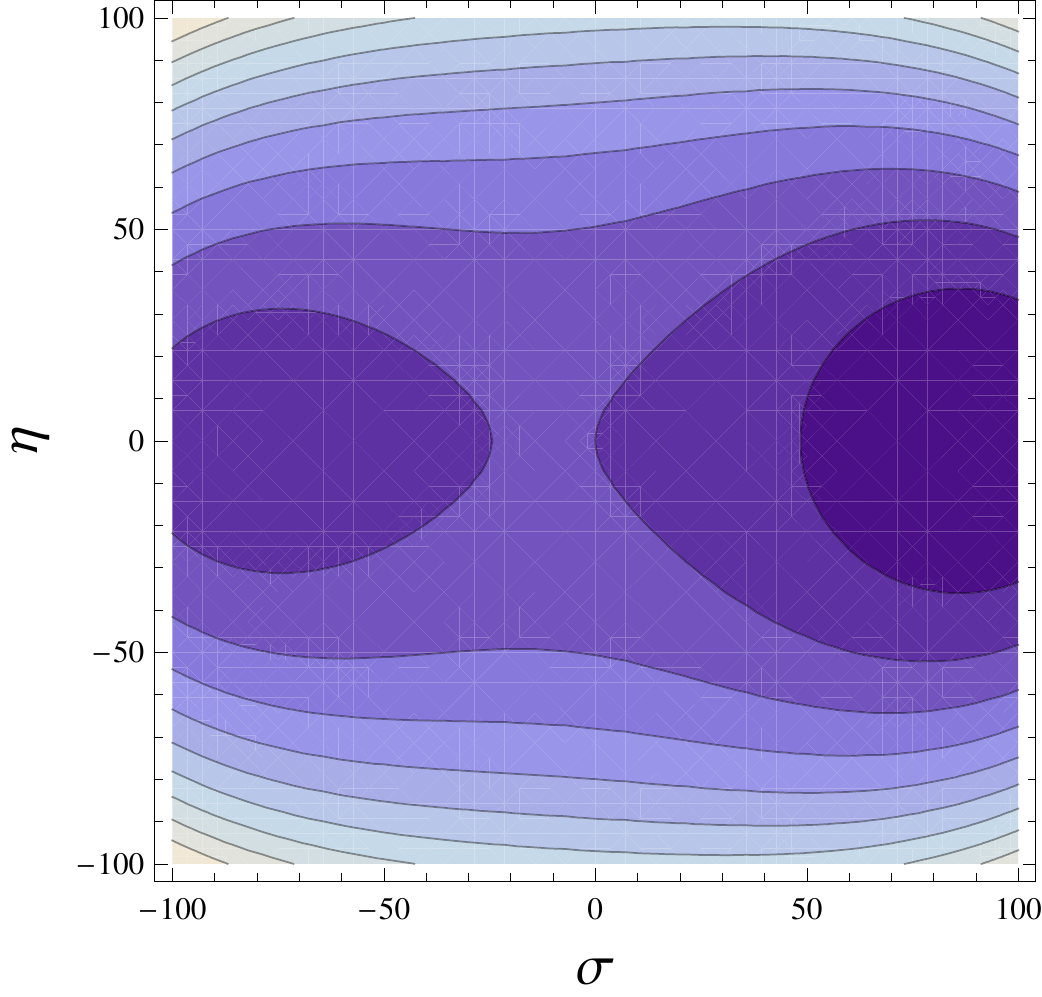,width=4.0cm,height=100pt}&
 \hspace{0.3cm} \epsfig{file=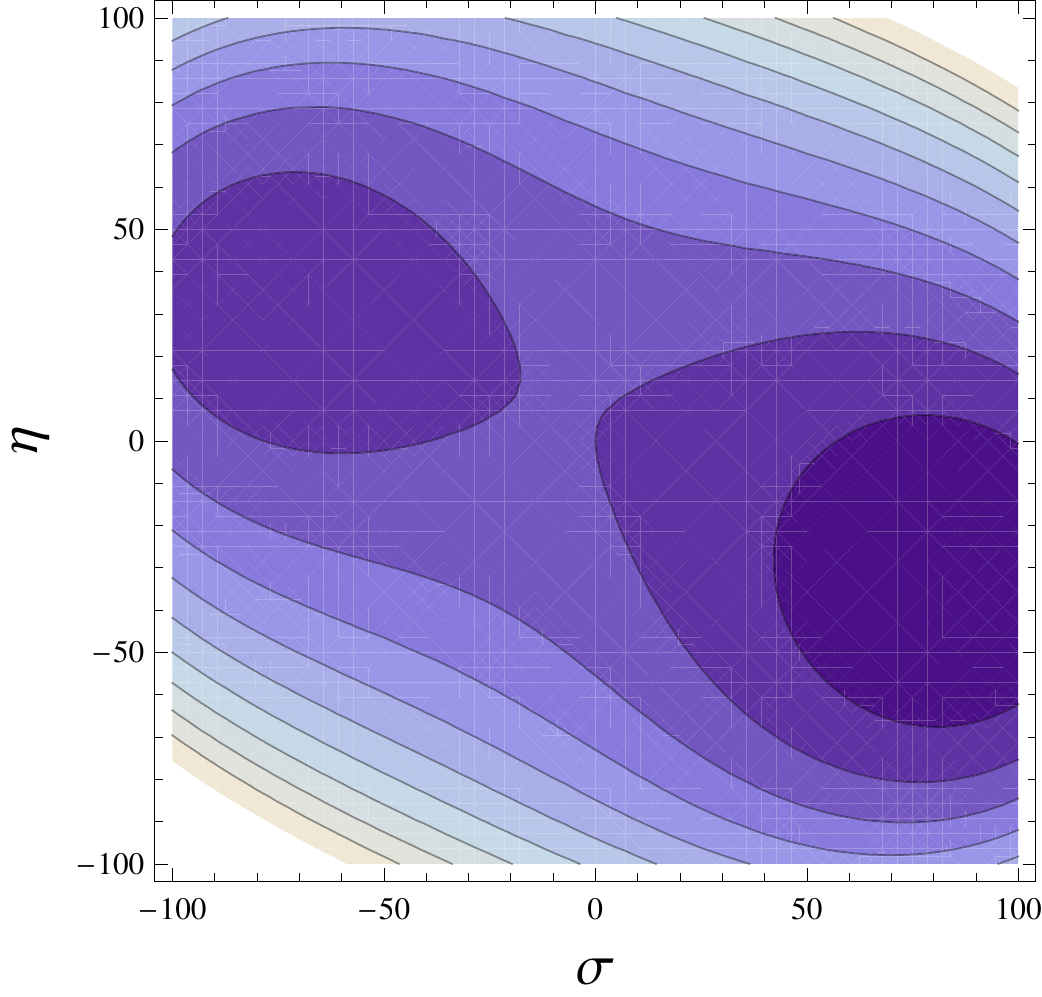,width=4.0cm,height=100pt}\\
\\
 (a) & (b) \\
\\
\\
\epsfig{file=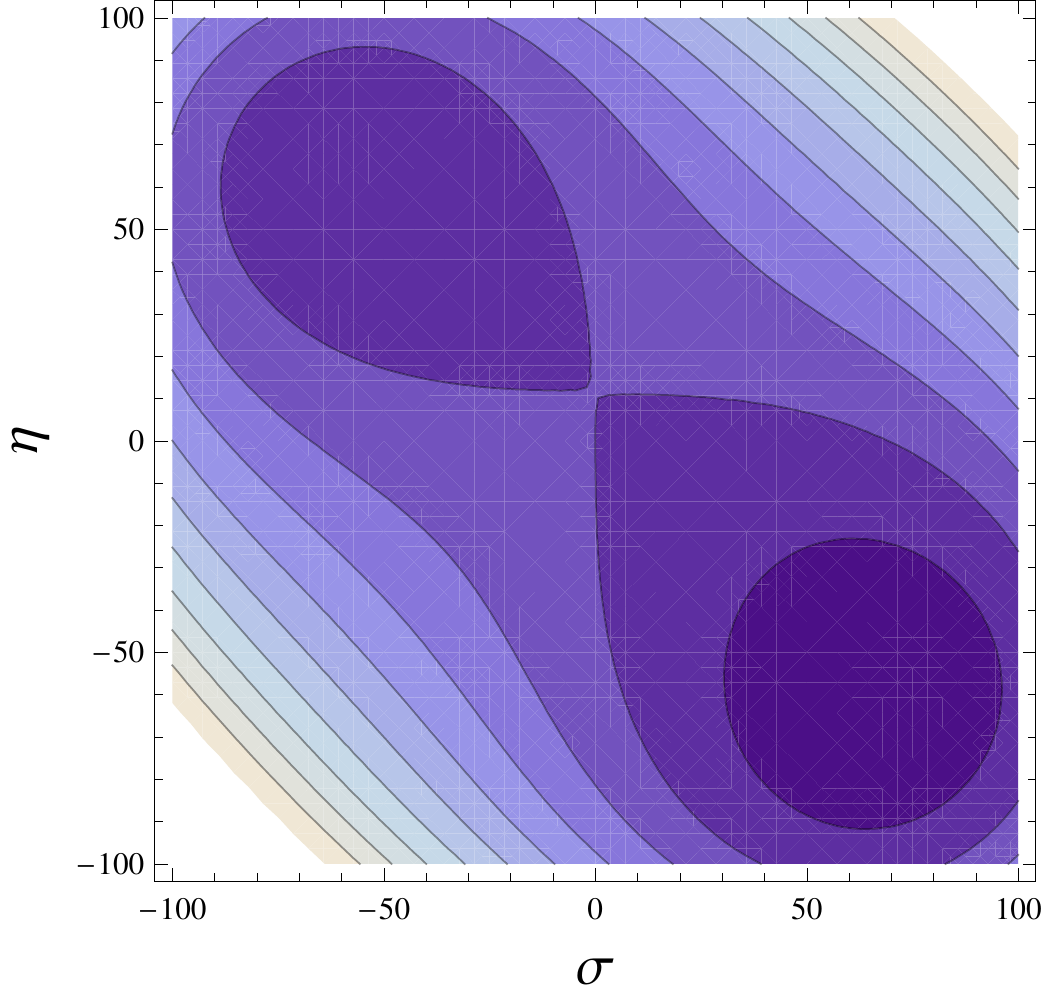,width=4.0cm,height=100pt}&
 \hspace{0.3cm} \epsfig{file=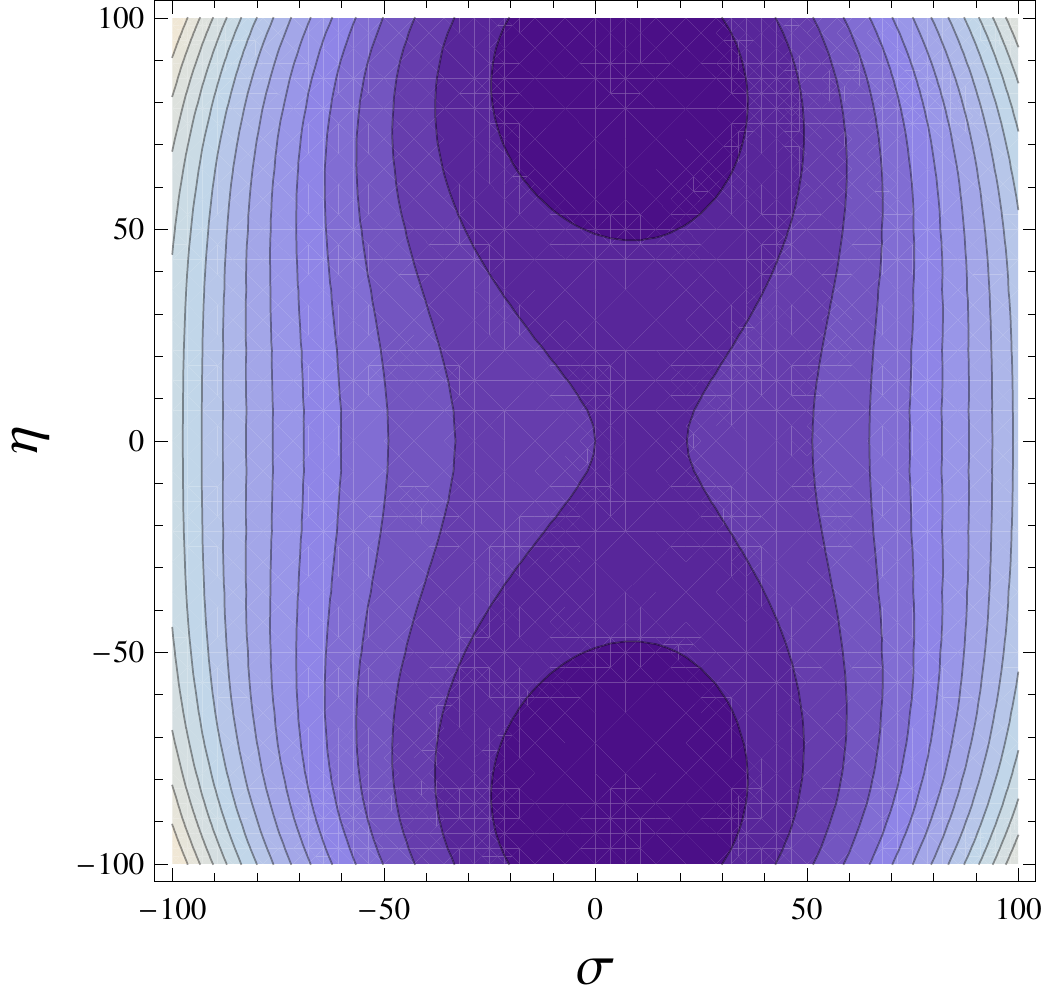,width=4.0cm,height=100pt}\\
\\
 (c)  & (d)
\end{tabular}
\end{center}
\caption{Contour plots of the effective potential in the vacuum.
(a) $\theta=0$; (b) $\theta=\pi/4$; (c) $\theta=\pi/2$;  (d) $\theta=\pi$.
The numerical values are in MeV.}
\vspace{0.5cm}
\label{fig:T=0}
\end{figure}

Following a mean field analysis, we take $\sigma = \sig + \sigma'$ and $\eta = \et + \eta'$, and
assume that the remaining condensates vanish. In this approximation, the effective mass for the
quarks is simply given by $M_{q}=g\sqrt{\sig^{2} + \et^{2}}$. The effective potential can then be
split into a classical piece and its fluctuation corrections, besides the contribution
coming from the quarks, i.e., $V_{eff} = V_{\chi}^{cl} + V_{\chi}^{fluct} + V_{q}$. The
classical contribution has the form
\ba \nonumber
V^{cl}_{\chi} &=& \frac{\lambda}{4} (\sig^2 - v_\theta^2)^2 - H\sig  \nonumber \\
&+&\frac{\lambda}{4}(\et^2 - u_\theta^2)^2 - c ~\sin\theta ~\sig\et  \nonumber \\
&+&\frac{\lambda}{2}\sig^2\et^2 - \frac{\lambda}{4}(v_\theta^4 + u_\theta^4)
\ea
where we have defined the combination of couplings $\lambda \equiv \lambda_1 + \lambda_2/2$,
and the following auxiliary quantities:
\be
v_\theta^2\equiv \frac{a+c ~\cos\theta}{\lambda}\quad ; \quad
u_\theta^2\equiv v_\theta^2 - \frac{2c}{\lambda} \cos\theta \; .
\ee
The fluctuation correction, up to quadratic terms, is straightforwardly worked out and
provides the means to fix the parameters of the model to meson masses and the pion
decay constant $f_{\pi}$ in the vacuum in the absence of CP violation. Concretely,
we impose that in the vacuum chiral symmetry is spontaneously broken and the expectation
value of the chiral condensate is given by $\sig=f_\pi=93~$MeV. The coefficient of the term
that breaks explicitly chiral symmetry is given by the PCAC relation $H=f_\pi m_\pi^2$. All
the other parameters are fixed to reproduce the mesons masses: $m_\sigma = 600~$MeV,
$m_\eta= 574~$MeV, $m_\pi=138~$MeV, and $m_{\alpha_0}=980~$MeV.

This completes the setup of the CP-odd linear sigma model to be used as an effective
theory in the investigation of the chiral transition in the presence of nontrivial CP
violating processes. In what follows we present our results for the effective potential
for representative values of the parameter $\theta$ and the temperature. Effects from a
strong magnetic background will be incorporated later.

\section{Results for the effective potential}

The effective potential we obtained in the previous section is a function of the condensates
of $\sigma$ and $\eta$, and of the CP violation parameter $\theta$. Below we analyze the
dependence of the effective potential on $\theta$ for different values of the temperature.
In the vacuum, i.e. in the zero temperature case, the potential has two minima in the $\sigma$
direction for $\theta=0$, as expected from usual (CP-even) linear sigma model. The contour
plot for the effective potential in the plane $\sig-\et$ for this situation is presented
in Fig. \ref{fig:T=0}(a). As we increase $\theta$, the pair of minima rotates, and for $\theta=\pi$ it
is almost in the $\et$ direction, as can be seen in Figs. \ref{fig:T=0}(b) and (c). Although this is not
a realistic (physical) case, since nonzero values of $\theta$ are not observed in the vacuum
of strong interactions, it is nevertheless interesting to study all the phase space to have a complete
map of the dependence on $\theta$.

\begin{figure}[!ht]
\begin{center}
\begin{tabular}{cc}
 \epsfig{file=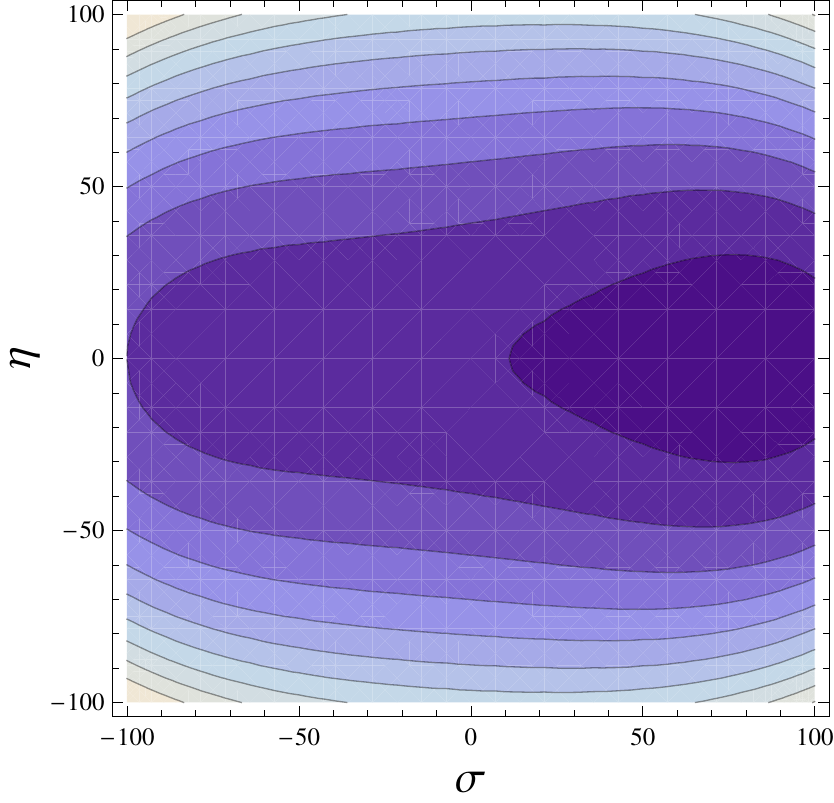,width=4.0cm,height=100pt}&
 \hspace{0.3cm} \epsfig{file=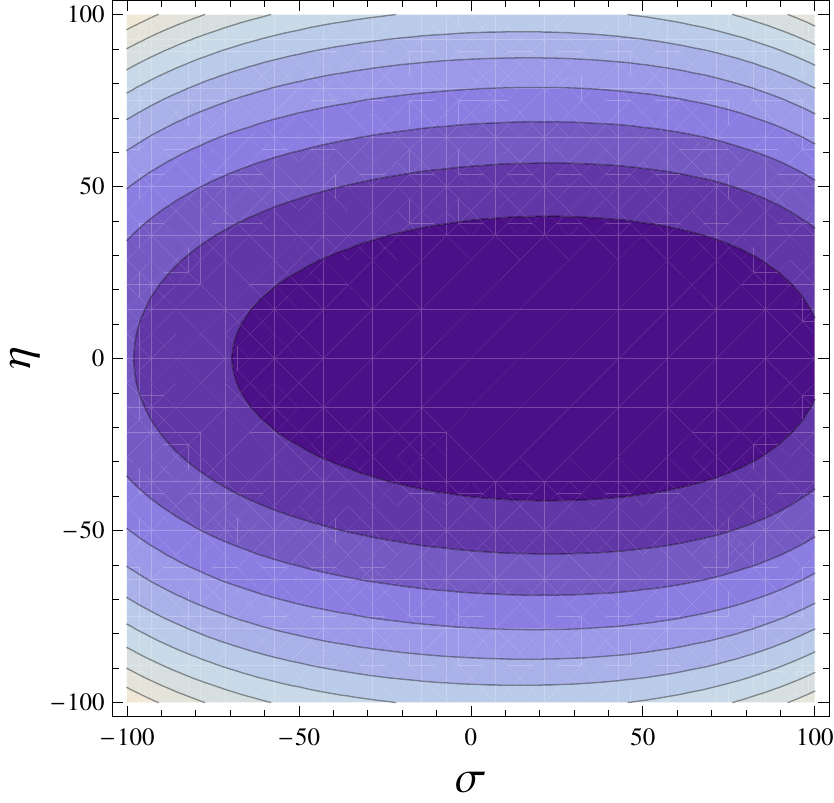,width=4.0cm,height=100pt}\\
\\
 (a) & (b) \\
\end{tabular}
\end{center}
\caption{Contour plots of the effective potential for $\theta=0$.
(a) $T=120~$MeV; (b) $T=160~$MeV.
The numerical values are in MeV.\\ \\}
\label{fig:theta=0_finitetemp}
\end{figure}

\begin{figure}[htbp]
\begin{center}
\includegraphics[width=9cm]{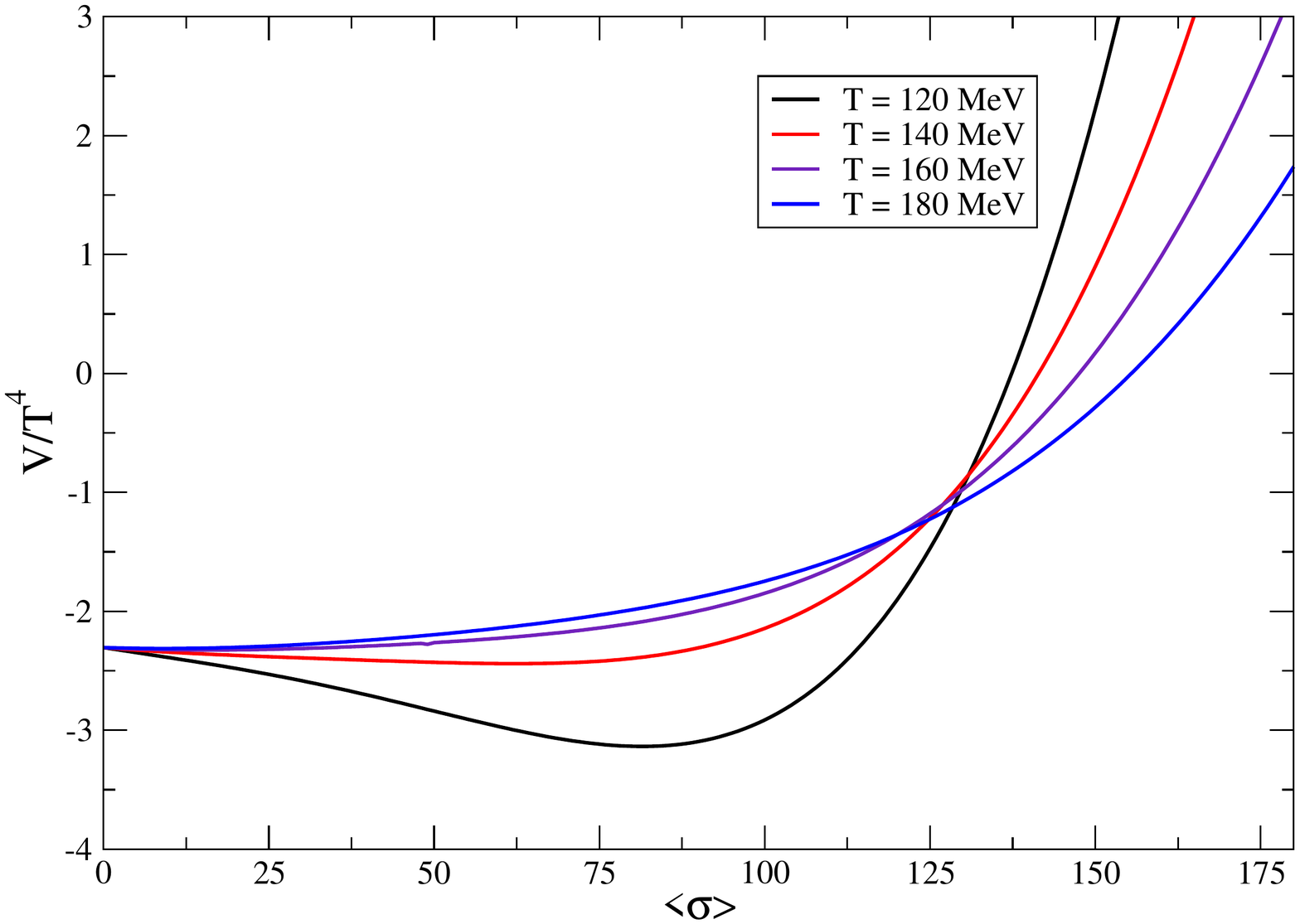}
\caption{Effective potential normalized by the temperature in the $\sig$ (in MeV)
direction at $\theta=0$ for several values of the temperature.}
\label{fig:sigmadirection}
\end{center}
\end{figure}

Keeping $\theta=0$ and increasing the temperature, the two minima remain in the $\sig$ direction,
as expected, and both minima approach to the origin, $\sig\approx 0$, so that for high enough
temperatures chiral symmetry is approximately restored (see Fig. \ref{fig:theta=0_finitetemp}).
Chiral symmetry is not completely restored due to the explicit breaking caused by the mass
term ($\sim H$) in the effective potential.
In Fig. \ref{fig:sigmadirection}, we plot the potential in the $\sig$ direction for $\theta=0$ and
different values of the temperature. As expected from the usual linear sigma model, the transition
to the phase with approximately restored chiral symmetry is a crossover.

\begin{figure}[!ht]
\begin{center}
\begin{tabular}{ccc}
 \epsfig{file=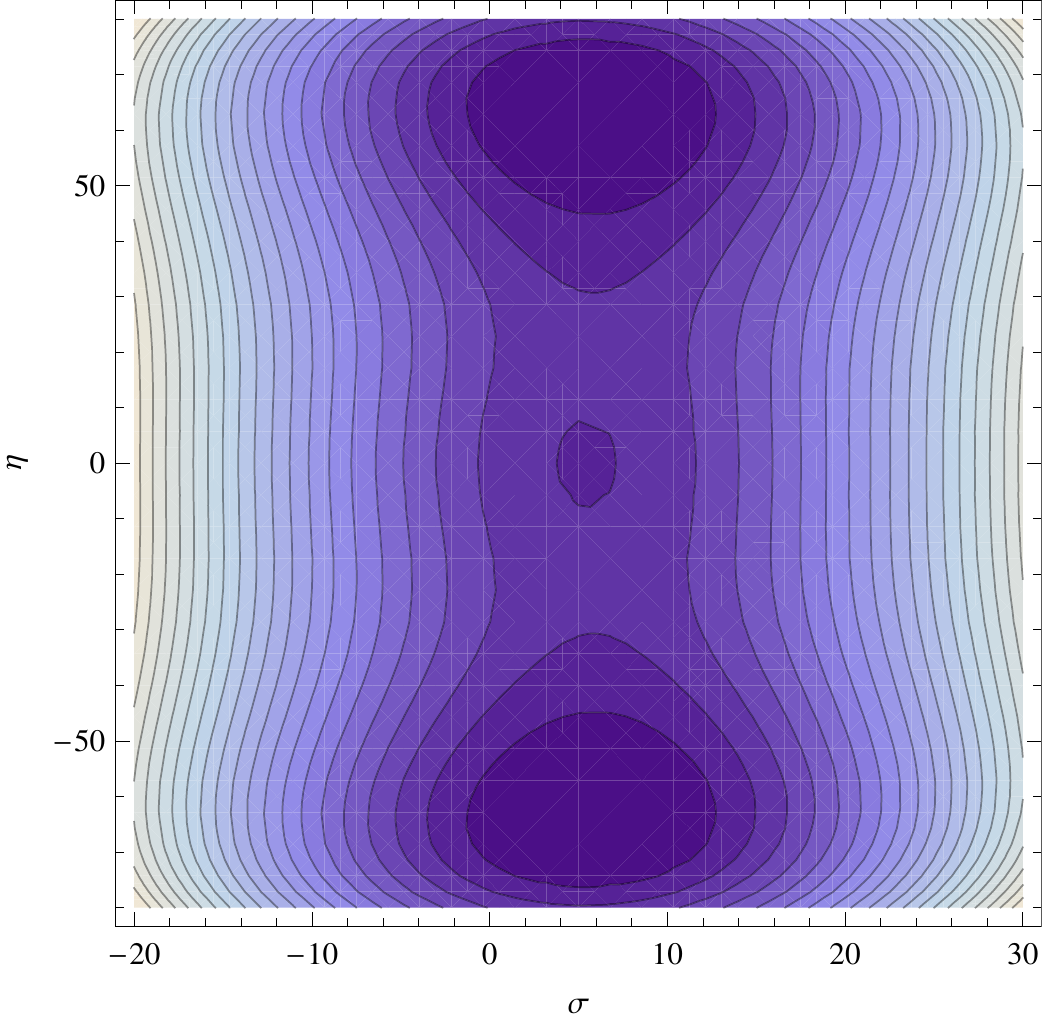,width=4.0cm,height=100pt}&
 \hspace{0.3cm} \epsfig{file=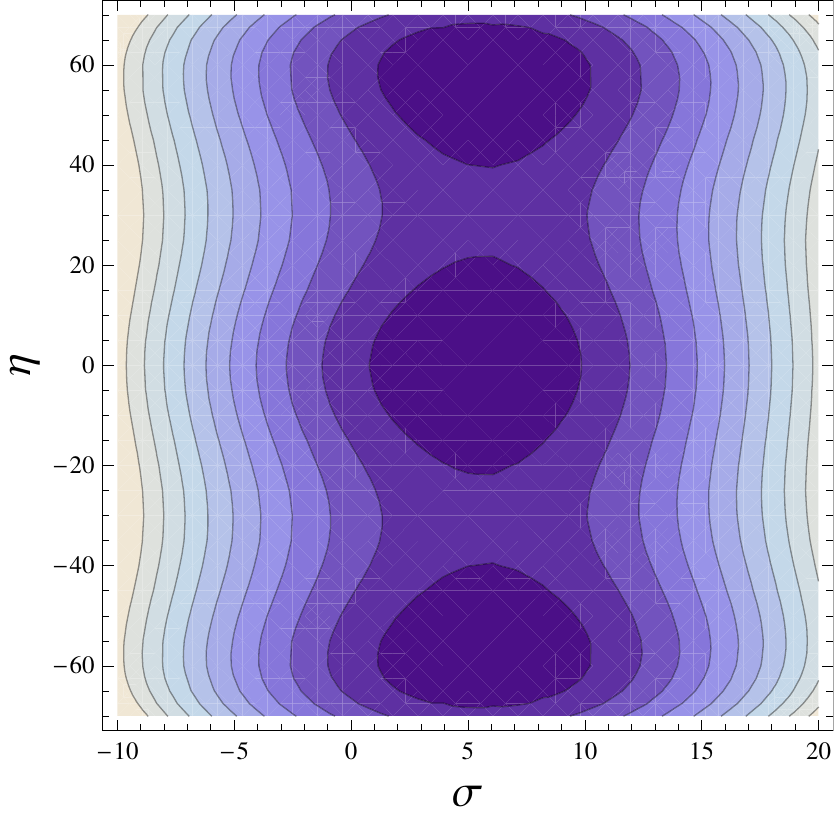,width=4.0cm,height=100pt}\\
\\
 (a) & (b) \\
\end{tabular}
\end{center}
\caption{Contour plots of the effective potential for $\theta=\pi$.
(a) $T=125~$MeV; (b) $T=128~$MeV.
The numerical values are in MeV.
}
\label{fig:theta=pi}
\end{figure}

For $\theta=\pi$, the degenerate minima are almost in the $\et$ direction, as it is shown in
Fig. \ref{fig:T=0}(d). Increasing the temperature, the minima move towards the center, indicating
a chiral symmetry restoration, as displayed in Fig. \ref{fig:theta=pi}. However, in this case, there is
a barrier between the global minimum and the new minimum that will become the true global minimum
at high temperature at $\eta=0$. In contrast to the previous case, this signals a first-order transition,
and the possibility of metastable CP-odd states.

\begin{figure}[htbp]
\begin{center}
\includegraphics[width=9cm]{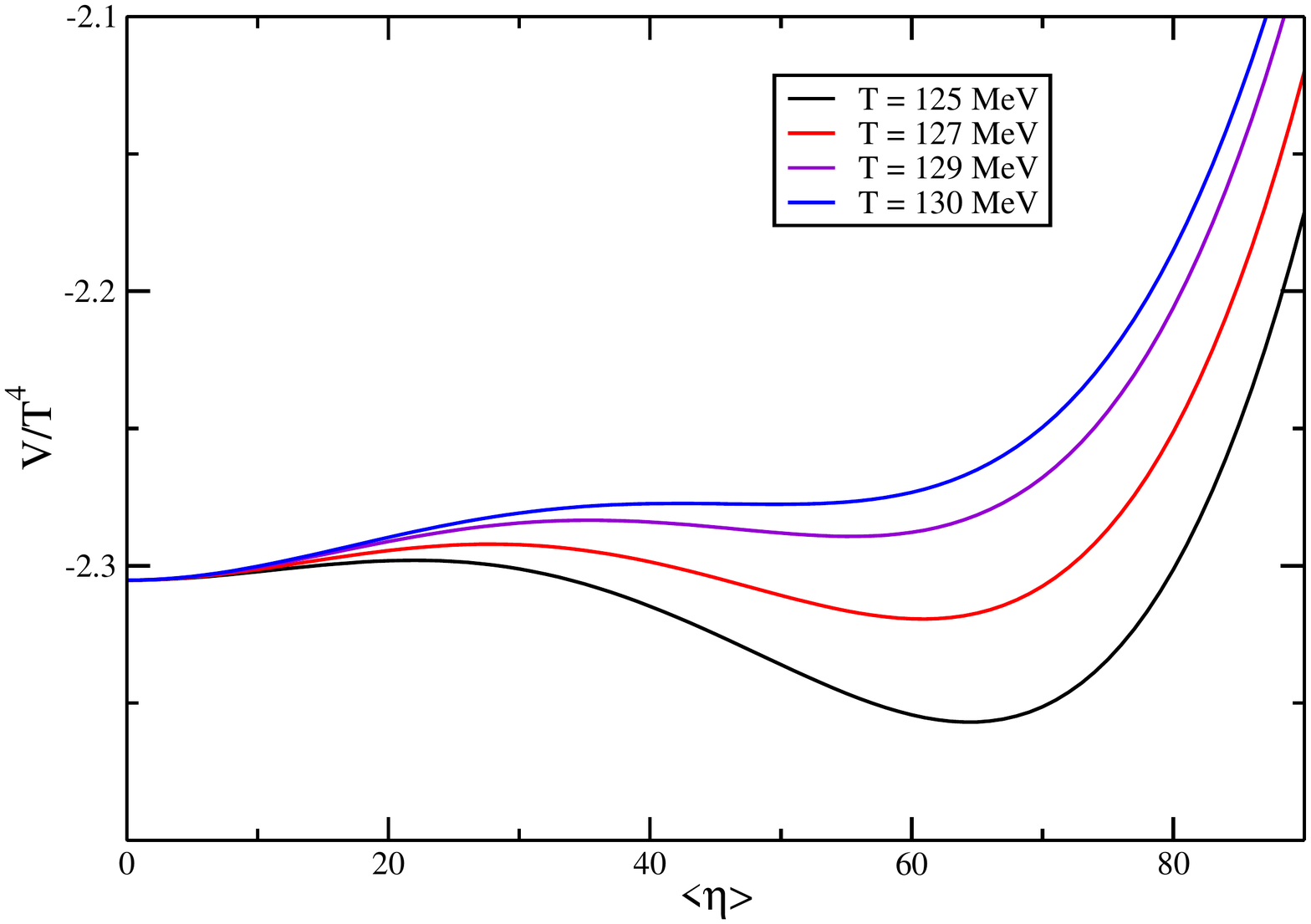}
\caption{Effective potential normalized by the temperature in the
$\et$ (in MeV) direction at $\theta=\pi$ for different values of the
temperature.} \label{fig:etadirection}
\end{center}
\end{figure}

To show more clearly the barrier, we display in Fig. \ref{fig:etadirection} the effective potential
at $\theta=\pi$ for different values of the temperature. One can notice the new minimum  emerging
at $\et=0$ for a temperature around $T=126~$MeV. In this case, the critical temperature is lower
than the critical temperature for the melting of the condensate $\sig$. Since the critical temperatures
for the melting of $\sig$ and $\et$ are different, three different phases are allowed in systems
with $\theta$ between zero and $\pi$: one in which both condensates are present, another one
where the $\et$ condensate vanishes, and a phase where both condensates vanish.

%
%
\begin{figure}[htbp]
\begin{center}
\includegraphics[width=9cm]{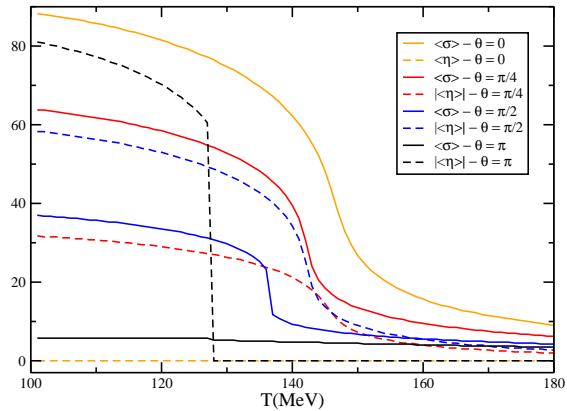}
\caption{Absolute value of the condensates (in MeV) as functions of
the temperature. As $\theta$ approaches $\pi$, the transition in the
$\et$ direction becomes stronger. Here full lines denote $\sig$ and
dotted lines $\et$. } \label{fig:minima}
\end{center}
\end{figure}
\begin{figure}[!ht]
\begin{center}
\begin{tabular}{ccc}
 \epsfig{file=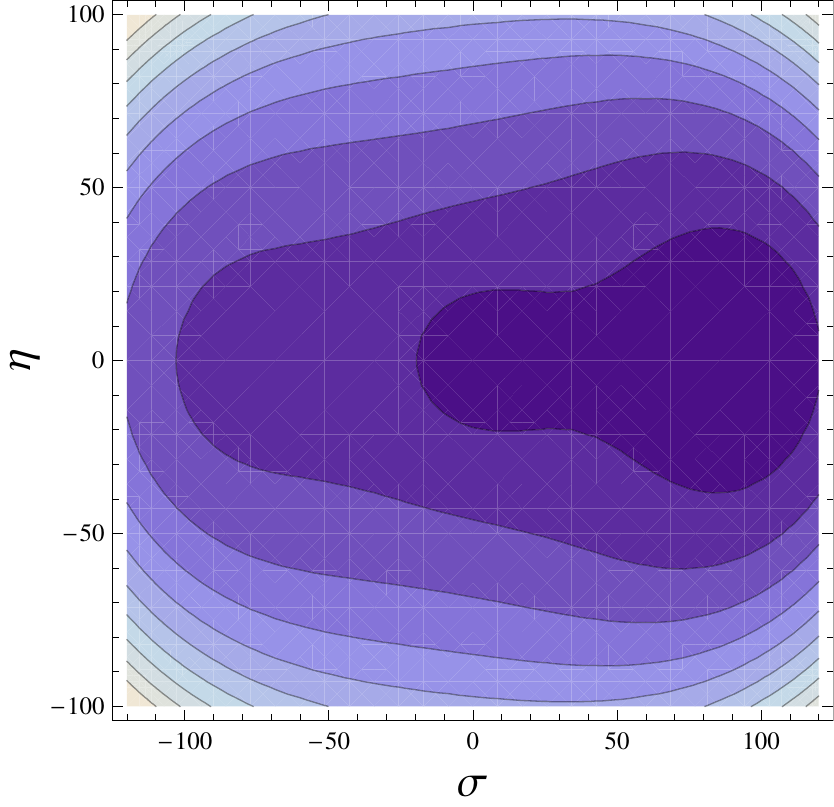,width=4.0cm,height=100pt}&
 \hspace{0.3cm}
 \epsfig{file=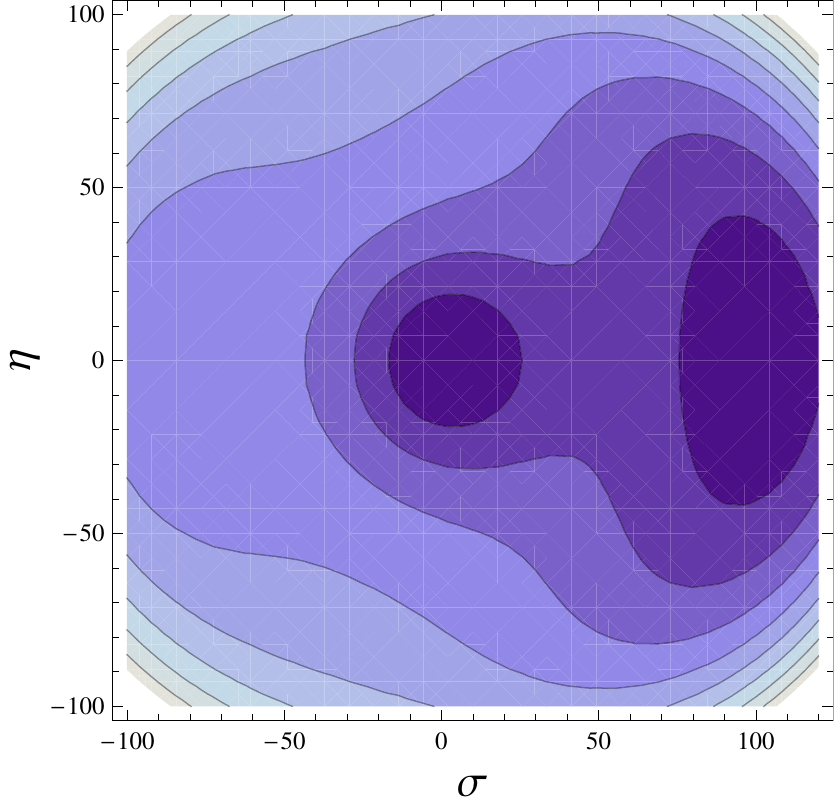,width=4.0cm,height=100pt}\\
\\
 (a) & (b) \\
\end{tabular}
\end{center}
\caption{Contour plots of the effective potential for $\theta=0$ and different values of $B$.
The temperature of each plot was chosen to be close to the respective critical temperature.
(a) $B=10m_\pi ^2$; (b) $B=20m_\pi ^2$. The numerical values are in MeV.}
\label{fig:mag_theta0}
\end{figure}

In Fig. \ref{fig:minima}, we plot the absolute value of the expectation values of $\sigma$ and $\eta$
as functions of the temperature for different values of $\theta$. Here, full lines stand for the $\sigma$
expectation value and dashed lines represent the condensate $\et$. Since their values are
``complementary'', the curves for different values of $\theta$ can be paired using the value of each
condensate at $T=100~$MeV: for $\theta=0$ the highest solid line corresponds to $\sig$ and
the lowest dashed line to $\et$, for $\theta=\pi/4$ the second to highest solid line corresponds
to $\sig$ and the second to lowest dashed line to $\et$, and so on. One can see that while $\sig$
decays smoothly for different values of $\theta$, the condensate $\et$ decays sharply when $\theta$
approaches $\pi$, indicating first-order phase transition. It is also clear in the figure that the critical
temperatures for the melting of the two condensates are different, allowing for the possibility of three
different phases, as mentioned above.

\section{Effects from a strong magnetic background}

Motivated by the possibility of observing the formation of CP-odd domains in high-energy heavy
ion collisions in the presence of strong magnetic fields for noncentral collisions, we can incorporate
the effect of a strong magnetic background on the effective potential for the CP-odd linear
sigma model, and study how the chiral transition and the condensates are modified. Assuming
that the system is now in the presence of a strong magnetic field background that is constant and
homogeneous, one can compute the modified effective potential following the procedure we
presented in detail in Ref. \cite{Fraga:2008qn}. In what follows, we simply sketch the method
and move to the discussion of the results.

\begin{figure}
\begin{center}
\begin{tabular}{ccc}
 \epsfig{file=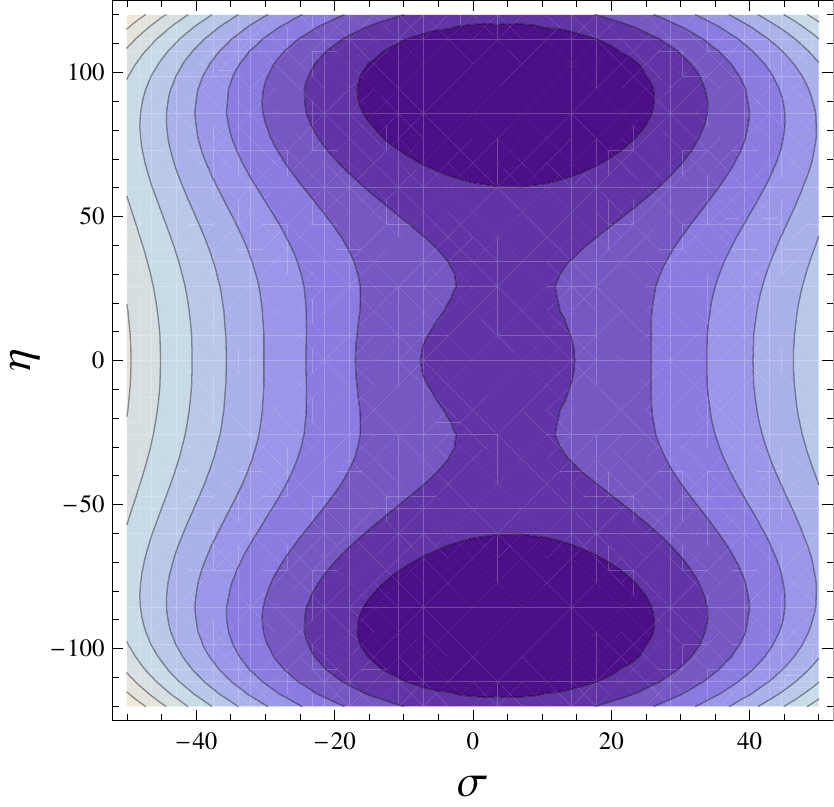,width=4.0cm,height=100pt}&
 \hspace{0.3cm} \epsfig{file=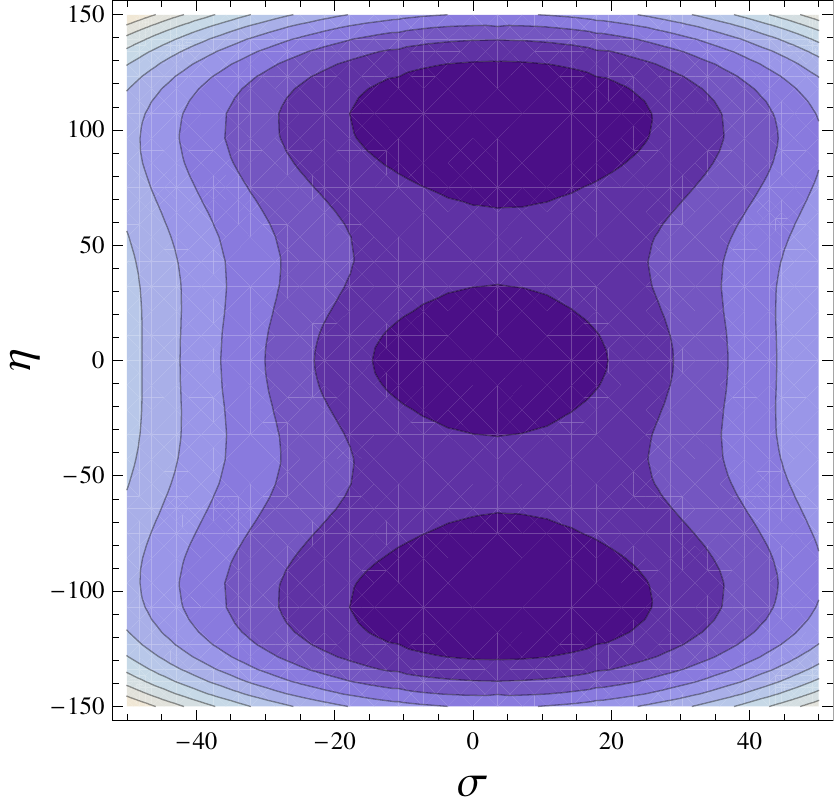,width=4.0cm,height=100pt}\\
\\
 (a) & (b) \\
\\
\\
\epsfig{file=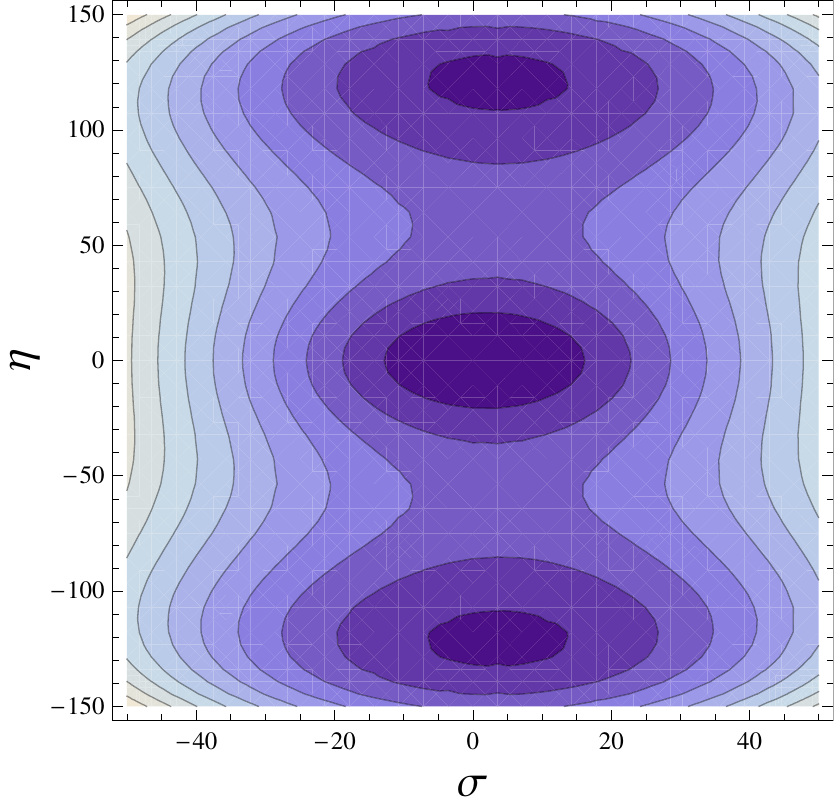,width=4.0cm,height=100pt}&
 \hspace{0.3cm} \epsfig{file=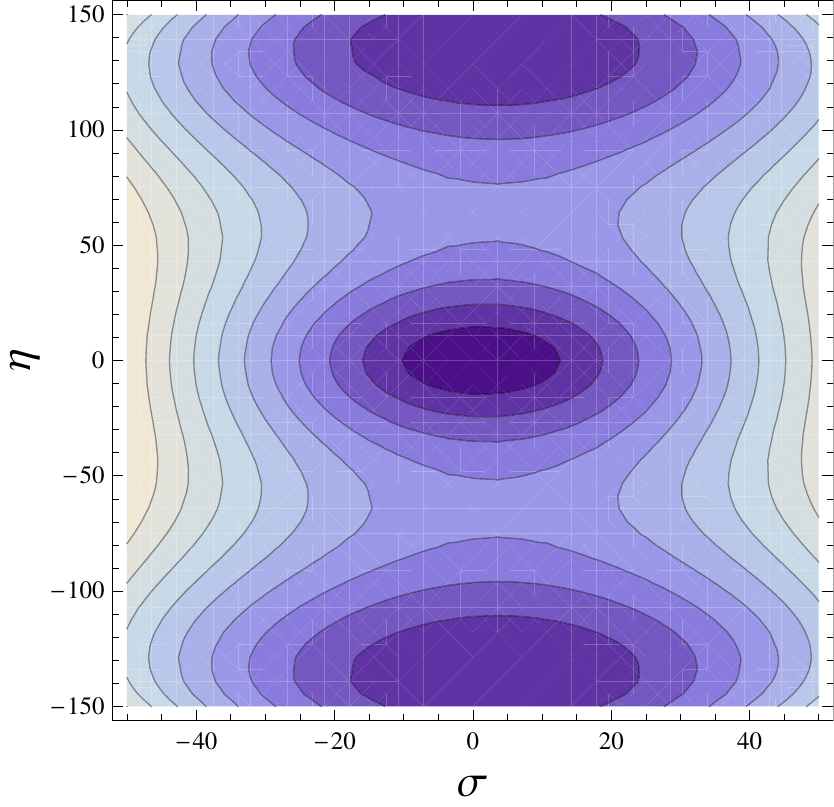,width=4.0cm,height=100pt}\\
\\
 (c)  & (d)
\end{tabular}
\end{center}
\caption{Contour plots of the effective potential for $\theta=\pi$ and different values of $B$.
The temperature was chosen to be $T=195~$MeV.
(a) $B=5m_\pi^ 2$; (b) $B=10m_\pi ^2$; (c) $B=15m_\pi ^2$;  (d) $B=20m_\pi ^2$.
The numerical values are in MeV.}
\label{fig:mag_thetapi}
\end{figure}
\begin{figure}
\begin{center}
\begin{tabular}{ccc}
\vspace*{-0.5cm}
 \epsfig{file=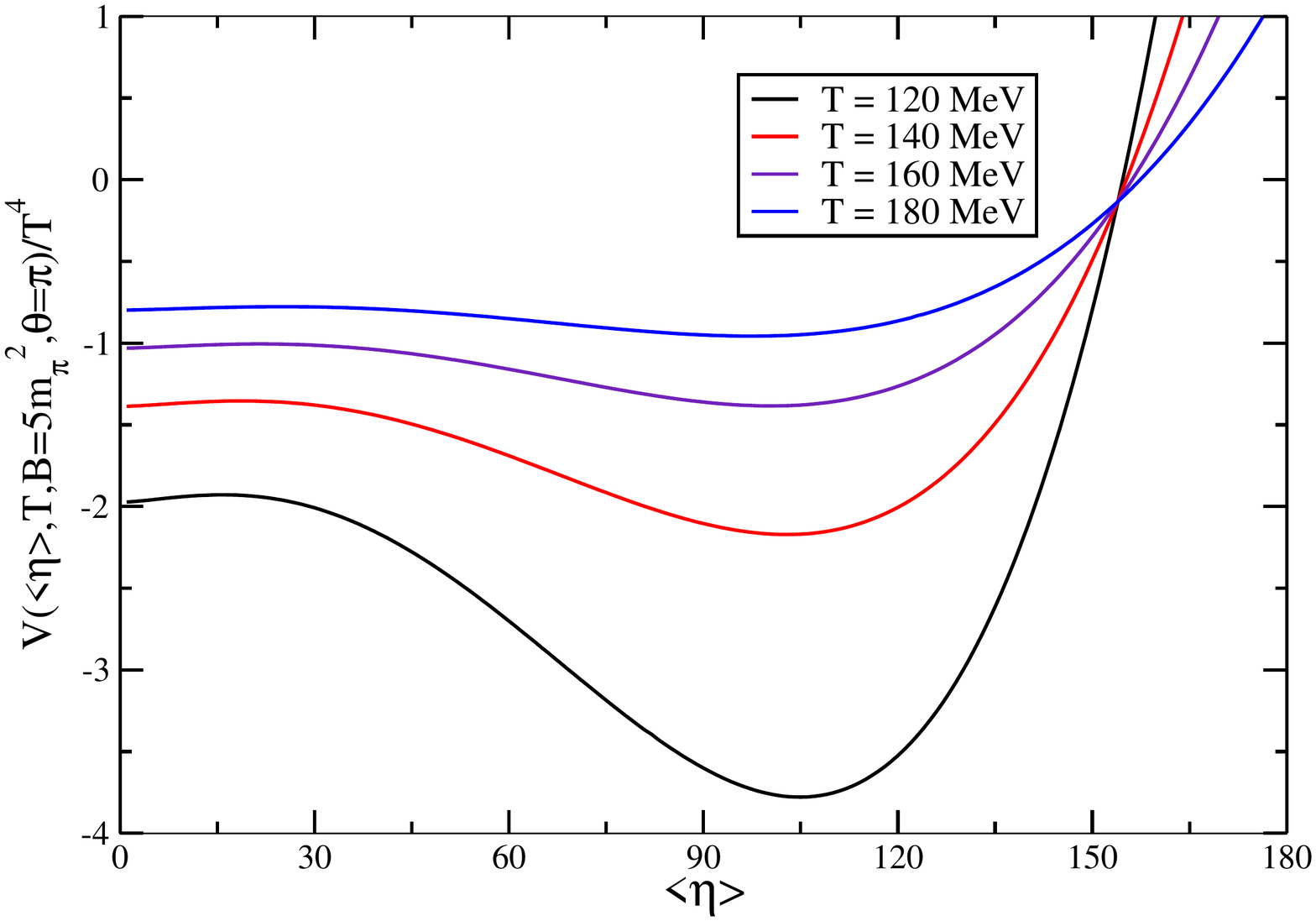,width=8cm}\\
\vspace{-0.6cm}
\epsfig{file=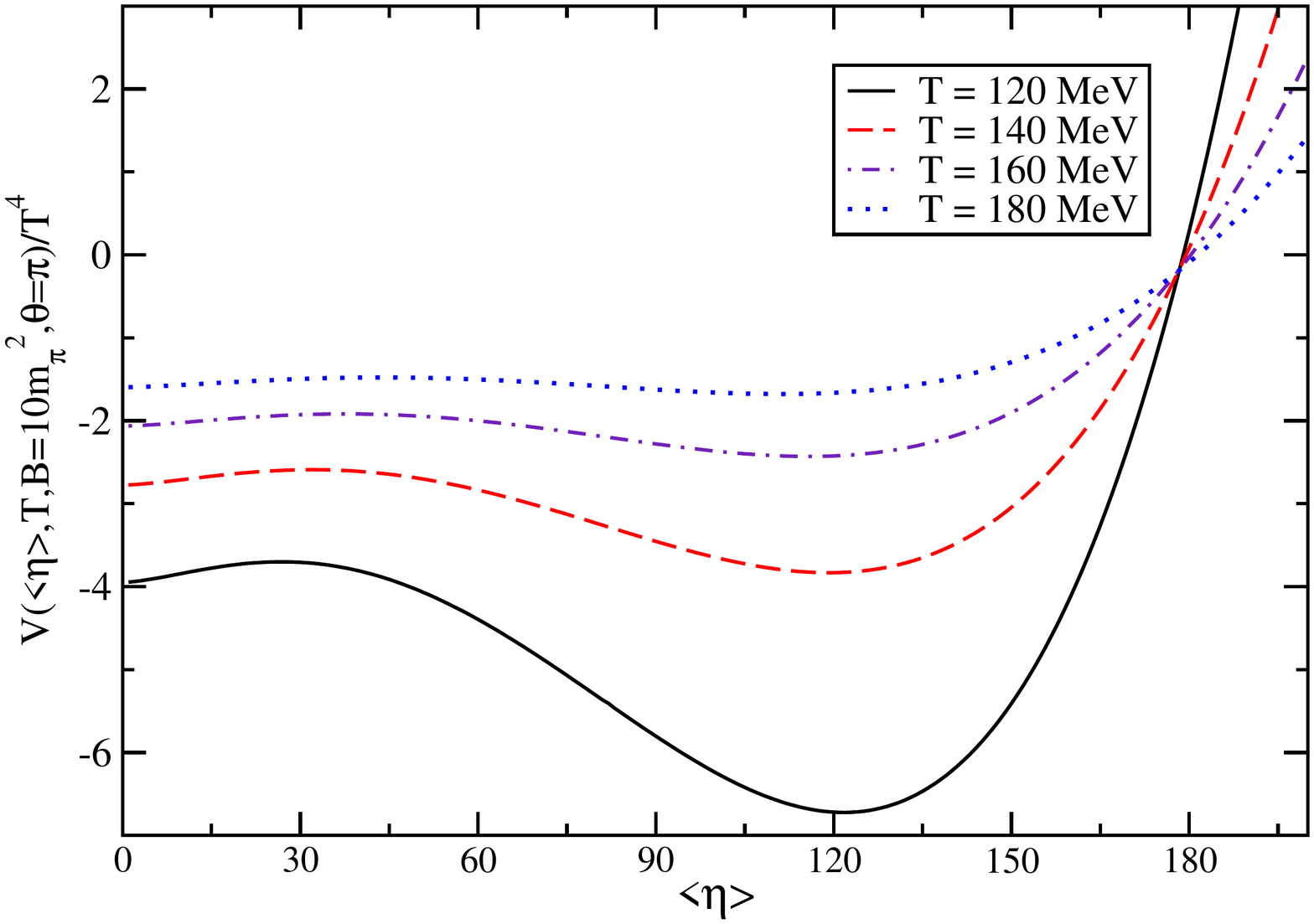,width=8cm}\\
\vspace{-0.6cm}
\epsfig{file=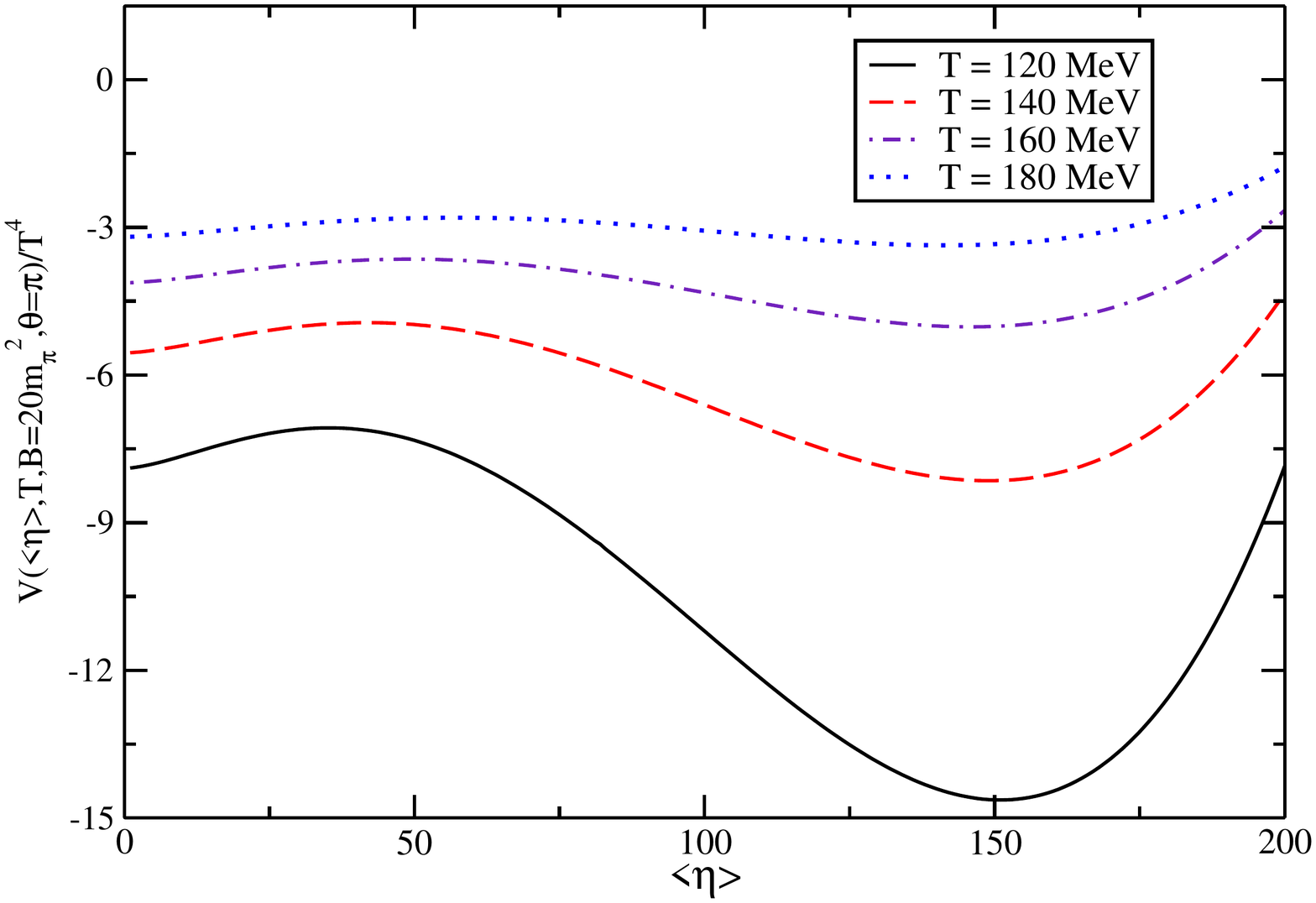,width=8cm}
\end{tabular}
\end{center}
\caption{Effective potential normalized by the temperature for $\theta=\pi$ in the $\et$ (in MeV)
direction. The plots, from the top to the bottom of the figure, correspond to $B=5m_\pi^2$,
$B=10m_\pi^2$, and $B=20m_\pi^2$.}
\label{fig:mag_slice}
\end{figure}

For definiteness, let us take the direction of the magnetic field as the $z$-direction, $\vec{B}=B \hat z$.
The effective potential can be generalized to this case by a simple redefinition of the dispersion relations
of the fields in the presence of $\vec{B}$, using the minimal coupling shift in the gradient and the field
equations of motion. For this purpose, it is convenient to choose the gauge such that
$A^{\mu}=(A^{0},\vec{A})=(0,-By,0,0)$. Decomposing the fields into their Fourier modes, one arrives
at eigenvalue equations which have the same form as the Schr\"odinger equation for a harmonic
oscillator potential, whose eigenmodes correspond to the well-known Landau levels. The latter
provide the new dispersion relations
\begin{eqnarray}
p_{0n}^2&=&p_z^2+m^2+(2n+1)|q|B \, ; \\
p_{0n}^2&=&p_z^2+m^2+(2n+1-\sigma)|q|B \, .
\end{eqnarray}
for scalars and fermions, respectively, $n$ being an integer, $q$ the electric charge,
and $\sigma$ the sign of the spin.
It is also straightforward to show that integrals over four momenta and thermal
sum-integrals acquire the following forms, respectively:
\begin{eqnarray}
&&\int \frac{d^4k}{(2\pi)^4} \mapsto \frac{|q|B}{2\pi}\sum_{n=0}^\infty
\int \frac{dk_0}{2\pi}\frac{dk_z}{2\pi} \, ,\\
T\sum_{\ell} &&\int \frac{d^3k}{(2\pi)^3} \mapsto \frac{|q|BT}{2\pi}\sum_\ell \sum_{n=0}^\infty
\int \frac{dk_z}{2\pi} \, ,
\end{eqnarray}
where $n$ represents the different Landau levels and $\ell$ stands for the Matsubara
frequency indices \cite{FTFT-books}.

In our effective model, the vacuum piece of the potential will be modified by the magnetic field
through the coupling of the field to charged pions. To one loop, and in the limit
of high $B$, $eB \gg m_{\pi}^{2}$, one obtains (ignoring contributions independent of the
condensates) \cite{Fraga:2008qn}
\begin{equation}
V_{\pi^+}^V+V_{\pi^-}^V=-\frac{2m_\pi^2 eB}{32\pi^2}\log 2 \, .
\label{Vpion}
\end{equation}

Thermal corrections are provided by pions and quarks. However, the pion thermal contribution
as well as part of the quark thermal contribution are exponentially suppressed for high magnetic
fields, as has been shown in Ref. \cite{Fraga:2008qn}. The only part of the quark thermal part
that contributes is
\be
V_q^T = -N_c \frac{eBT^2}{2\pi^2} \left[\int_{-\infty}^{+\infty} dx~
\ln\left( 1+e^{-\sqrt{x^2 +M_q^2/T^2}}\right)\right] \; ,
\label{Vquark}
\ee
where $N_{c}=3$ is the number of colors. Therefore, the effective potential computed
in the previous section is corrected by the contributions in (\ref{Vpion}) and (\ref{Vquark})
in the presence of a strong homogeneous magnetic background.

For $\theta=0$, i.e. in the case of the CP-even linear sigma model, the situation is the same
as the one we investigated in Ref. \cite{Fraga:2008qn}. Initially, there are two minima in the
$\sig$ direction, one global and one local. As the temperature increases, they move towards
the center, approximately restoring chiral symmetry (not fully restored since there is an explicit
breaking term in the effective potential, due to the nonzero quark masses). For large enough values
of the magnetic background, the nature of the phase transition is altered, becoming of first order
instead of a crossover, in accordance with the findings of Ref. \cite{Fraga:2008qn}.
In Fig. \ref{fig:mag_theta0} we show contour plots of the full effective potential at $\theta=0$
for different values of the magnetic field. In each plot we display the potential close to the critical
temperature, so that the case of $B=10m_\pi^2$ is plotted at $T=120~$MeV, while the case
of $B=20m_\pi^2$ is plotted at $T=100~$MeV. For $B=5m_\pi^2$, the barrier is quite small
and the critical temperature higher, as discussed in Ref. \cite{Fraga:2008qn}.

Augmenting the value of $\theta$ to $\pi$, we can study the behavior of the $\eta$ condensate.
As in the case in the absence of a magnetic background, this condensate undergoes a first-order
phase transition. Moreover, in the presence of a strong magnetic field, its critical temperature is
strongly affected. Analogously to our findings in Ref. \cite{Fraga:2008qn} for the CP-even case,
a field  $B=5m_\pi^2$ increases the critical temperature, but now for the $\eta$ condensate.
The behavior of the critical temperature for stronger values of the field is analogous to the one
of $\sig$, i.e. it drops considerably.
In Fig. \ref{fig:mag_thetapi} we plot the effective potential at temperature $T=195~$ MeV.
For $B=5m_\pi^2$, this temperature is not enough to take the system to the critical region.
On the other hand, for $B=10m_\pi^2$ this temperature is close to the critical temperature,
and for $B=20m_\pi^2$  it is larger than the critical value. This can also be clearly seen in
Fig. \ref{fig:mag_slice}, where we display slices of the effective potential in the $\eta$ direction
for different values of magnetic field and temperature. 

Besides modifying the value of $T_c$,
the presence of a strong magnetic background deepens the absolute minimum and favors
a first-order transition.

\section{Conclusions and outlook}

The possibility of probing the vacuum structure of QCD in high-energy heavy ion
collisions, bringing some light to the understanding of the strong CP problem, is a very
exciting prospect. The consideration of the theoretical description of the environment that
might be produced under the appropriate experimental conditions, especially the major
role played by the presence of a strong magnetic background, has already brought up
very interesting phenomena, such as the chiral magnetic effect \cite{Kharzeev:2007jp}
and the possibility of converting the nature of the chiral transition from a crossover to
a first-order phase transition \cite{Fraga:2008qn}.

The complete physical scenario in heavy ion collisions is rather
complicated \cite{Kharzeev:1998kz,Kharzeev:1999cz,Kharzeev:2007jp}, so that
it is more prudent to consider, theoretically, the role of each relevant ingredient
separately, aiming at a more complete picture at the end. Previously, we have
investigated the effect of a strong magnetic background on the nature of the
chiral transition \cite{Fraga:2008qn}, obtaining remarkable effects as mentioned
above, and opening a new line of possibilities in the study of the phase structure
of strong interactions. In this paper, we investigated how the chiral transition,
more specifically how the effective potential written in terms of the condensates
$\sig$ and $\et$, is altered when one includes the possibility of strong CP violation
and a strong magnetic background. For this purpose we built a CP-odd extension
of the linear sigma model coupled with two flavors of quarks at finite temperature
and zero density. The inclusion of magnetic effects is implemented via a redefinition
of the dispersion relations.

We found that, in the absence of magnetic fields, the $\sigma$ condensate behaves
pretty much like it does in the usual CP-even linear sigma model, changing from
a phase where chiral symmetry is broken ($\sig\neq 0$) to a phase where it is approximately
restored ($\sig\approx 0$) via a crossover as the temperature is increased.
For non-vanishing values of the parameter $\theta$, a condensate of the field  $\eta$ builds up
for low temperatures, breaking CP. This condensate is trapped by a barrier in the effective
potential, indicating the presence of metastable states that violate CP, which is a relevant
point for the possibility of observing CP-odd domains in high-energy heavy ion collisions.
We should remark that these metastable states were not found in a detailed study of the
phase diagram defined in terms of the strength of the 't Hooft determinant, quark masses,
temperature and chemical potential that has been performed for $\theta =\pi$ within the
two-flavor NJL model \cite{Boer:2008ct}. The reason for this discrepancy has its grounds
in the difference in physical content of the two effective theories, and is currently under
investigation \cite{metastable}.

The main effect of a high magnetic background on the chiral transition is turning it into a
first-order transition, even in the case of the CP-odd linear sigma model, deepening the
absolute minimum of the effective potential. The behavior of the critical temperature is also
affected in a non-trivial way, first going up and then dropping for larger values of the
magnetic field, as was also observed in Ref. \cite{Fraga:2008qn}.

In order to address the physical scenario in the case of noncentral heavy ion collisions
at RHIC and the LHC, one has to incorporate the effects from nonzero chiral and baryonic
chemical potentials \cite{Kharzeev:2007jp}. It is also crucial to describe the nonequilibrium
dynamics of the formation of the condensates, which will provide the relevant time scales
and give information on the actual possibility of measuring effects coming from the formation
of CP-odd domains. This issues will be considered in a future publication \cite{future}.

\section*{Acknowledgments}
We thank J. Boomsma, S. A. Dias, A. Schmitt, and especially D. Boer for fruitful discussions.
This work was partially supported by CAPES, CNPq, FAPERJ and FUJB/UFRJ.


\end{document}